\documentclass[submission, Proceedings]{SciPost}

\hypersetup{
    colorlinks,
    linkcolor={red!50!black},
    citecolor={blue!50!black},
    urlcolor={blue!80!black}
}

\usepackage[bitstream-charter]{mathdesign}
\DeclareSymbolFont{usualmathcal}{OMS}{cmsy}{m}{n}
\DeclareSymbolFontAlphabet{\mathcal}{usualmathcal}

\newcommand{\kkmchh}{{\tt KKMChh}}
\newcommand{\kkhhfoam}{{\tt KKhhFoam}}

\newcommand{\Mcal}{{\cal M}}

\newcommand{\Ocal}{{\cal O}}

\newcommand{\Mmf}{\mathfrak{M}}
\begin{document}
\begin{center}{\Large \textbf{
New Developments in KKMChh: Quark-Level Exponentiated Radiative Corrections and Semi-analytical Results\\
}}\end{center}

\begin{center}
S.A. Yost\textsuperscript{1$\star$},
M. Dittrich\textsuperscript{2}
S. Jadach\textsuperscript{3},
B.F.L. Ward\textsuperscript{4} and
Z. W{\c a}s\textsuperscript{5}
\end{center}

\begin{center}
{\bf 1} The Citadel, Charleston, South Carolina, USA
\\
{\bf 2} University of Florida, Gainesville, Florida, USA
\\
{\bf 3} Institute of Nuclear Physics, Polish Academy of Sciences,
Kraków, Poland
\\
{\bf 4} Baylor University, Waco, Texas, USA
\\
\textsuperscript{$\star$} scott.yost@citadel.edu
\end{center}

\begin{center}
\today
\end{center}


\definecolor{palegray}{gray}{0.95}
\begin{center}
\colorbox{palegray}{
  \begin{tabular}{rr}
  \begin{minipage}{0.1\textwidth}
    \includegraphics[width=35mm]{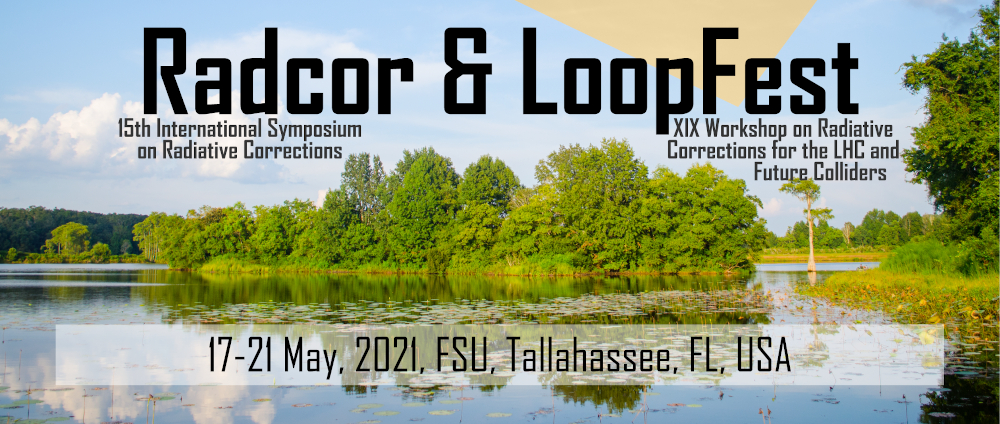}
  \end{minipage}
  &
  \begin{minipage}{0.85\textwidth}
    \begin{center}
    {\it 15th International Symposium on Radiative Corrections: \\Applications of Quantum Field Theory to Phenomenology,}\\
    {\it FSU, Tallahassee, FL, USA, 17-21 May 2021} \\
    \doi{10.21468/SciPostPhysProc.?}\\
    \end{center}
  \end{minipage}
\end{tabular}
}
\end{center}

\section*{Abstract}
{\bf
We describe a new semi-analytical program, KKhhFoam, which provides a simplified framework for testing the amplitude-level exponentiation scheme (CEEX) of the full KKMChh program in the semi-soft limit. The structure of the KKhhFoam integrand is also helpful for elucidating the structure of CEEX. We also discuss the representation of ISR in KKMChh and compare the ISR added by KKMChh to the effect of switching to a QED-corrected PDF, at the individual quark level, and suggest a new approach to running KKMChh with QED-corrected PDFs.
}


\section{Introduction}
\label{sec:intro}

KKMChh\cite{KKMChh-PhysRevD94,KKMChh-PhysRevD99,jadach2021ifi} is an adaptation of the LEP-era Monte Carlo event generator KKMC\cite{Jadach:1999vf} to the hadronic Drell-Yan process including exponentiated multi-photon effects at the quark level process
\begin{equation}
    q{\overline q}\rightarrow Z/\gamma^* \rightarrow l{\overline l} + n\gamma 
\end{equation}
into leptons including exact $\Ocal(\alpha)$ and $\Ocal(\alpha^2 L)$ QED initial state radiation (ISR), final state radiation (FSR), and initial-final interference (IFI), where $L$ is a ``big logarithm'' appropriate to each type of radiation. KKMChh is one of several currently available programs adding photonic and electroweak (EW) corrections to hadronic scattering. Other programs with comparable capabilities include MC-SANC\cite{MC-SANC}, POWHEG-EW\cite{POWHEG-EW}, HORACE\cite{horace1,horace2,horace3,horace4}, ZGRAD\cite{ZGRAD,ZGRAD2}, and RADY\cite{Dittmaier2020}, some of which are compared in an LHC EW benchmark study \cite{2016benchmark}.

Two types of soft-photon exponentiation are supported in KKMChh: exclusive exponentiation (EEX), which is YFS-style exponentiation\cite{Yennie:1961ad} at the cross section level, and coherent exclusive exponentiation (CEEX) \cite{Jadach:1998jb,Jadach:2000ir}, which is implemented at the spin-amplitude level.\cite{Jadach:1998wp} Only CEEX mode supports IFI. Order $\alpha$ electroweak matrix element corrections are included via an independent DIZET6.45\cite{Arbuzov2006,Arbuzov2020} module that tabulates EW form factors before a KKMChh run. Originally developed as a mixture of FORTRAN and C++, KKMChh has been recently reprogrammed entirely in C++. This will facilitate compilation on a broader range of platforms and integration with modern parton showers. Upgrading the original HERWIG6.5\cite{HERWIG6} interface in KKMChh to HERWIG7\cite{HERWIG7} is a work in progress.

We will focus here on a new semi-analytical program KKhhFoam developed for testing the soft-photon limit of KKMChh. This program is also useful to elucidate the CEEX exponentiation structure of KKMChh in a simplified and more intuitive context. We will also discuss the ISR implementation in KKMChh, and discuss its relation to parton distribution functions (PDFs) which either include or neglect the effect of QED evolution.

\section{KKhhFoam: The Semi-Soft Approximation}
\label{sec:KKhhFoam}

KKhhFoam is a hadronic adaptation of the semi-analytical program KKFoam\cite{Jadach:2018lwm} for $e^+e^-$ scattering, which is in turn an adaptation of KKsem\cite{Jadach:2000ir}, a predecessor which included ISR and FSR only. Both KKFoam and KKhhFoam include exponentiated IFI as well. These programs adopt a semi-soft approximation where the loss of momentum to ISR is included in the matrix element but hard photon corrections to the radiation are neglected. If a cutoff on the maximum radiated photon energy is included in both KKhhFoam and KKMChh, the programs should agree for sufficiently inclusive observables, providing a way to compare KKMChh to a much simpler implementation of CEEX exponentiation. This simpler implementation is also easier to understand than the full KKMChh implementation, and is useful to elucidate the structure of CEEX exponentiation. 

We will see that ISR, FSR, and IFI are described by separate radiator functions -- in fact two of them in the case of IFI. 
Following the development of ref.\ \cite{Jadach:2018lwm}, the structure of the CEEX matrix element, neglecting non-soft parts, may be expressed as 
\begin{equation}
    \sigma(s) = \frac{1}{{\rm flux}(s)}\sum_{n=0}^\infty \frac{1}{n!}\int d\tau_{n+2} \Mmf^{\mu_1,\cdots,\mu_n}(k_1,\cdots,k_n)
    \left[\Mmf_{\mu_1,\cdots,\mu_n}\right]^*
\end{equation}
where the $k_i$ are $n$ photon momenta and the phase space includes also the quark and anti-quark momenta $p_1, p_2$. The final state fermion and anti-fermion momenta $q_1, q_2$ are constrained by $q_1 + q_2 = p_1 + p_2 - \sum_{i=1}^n k_i$. The spin amplitudes have the form 
\begin{equation}
    \Mmf^{\mu_1,\cdots,\mu_n}(k_1,\cdots,k_n) = \sum_{V=\gamma,Z} e^{\alpha(B_4 + \Delta B_4^V)} \sum_{\{I,F\}}\prod_{i\in I} J_I^{\mu_i}(k_i) \prod_{f\in F}J_F^{\mu_f}(k_f) \Mcal_V^{(0)}\left(p_1+p_2+\sum_{j\in I}k_j\right)
\end{equation}
with the sum over $\{I,F\}$ a sum over all partitions of the $n$ photons into initial and final state sets $I, F$, and initial and final state currents 
\begin{equation}
    J_I^\mu(k) = \frac{Q_I e}{4\pi^{3/2}}\left(\frac{p_1^\mu}{p_1\cdot k}-\frac{p_2^\mu}{p_2\cdot k}\right), \quad
    J_F^\mu(k) = \frac{Q_F e}{4\pi^{3/2}}\left(\frac{q_1^\mu}{q_1\cdot k}-\frac{q_2^\mu}{q_2\cdot k}\right),
\end{equation}
where $Q_I, Q_F$ are the quark and lepton charges. The YFS virtual form factor is 
\begin{equation}
    B_4 = Q_I^2 B_2(p_1,q_1) + Q_F^2 B_2(q_1,q_2) + Q_I Q_F\left[B_2(p_1,q_1)+B_2(p_2,q_2) - B_2(p_1,q_2) - B_2(p_2,q_1)\right]
\end{equation}
with 
\begin{equation}
    B_2(p,q) = \frac{i}{(2\pi)^3}\int \frac{d^4 k}{k^2-m_\gamma^2+i\epsilon} \left(\frac{2p+k}{k^2+2p\cdot k+i\epsilon} + \frac{2q-k}{k^2-2q\cdot k+i\epsilon}\right).
\end{equation}
There is also a resonant virtual form factor $\Delta B_4^V$ which resums logarithms in $\Gamma_Z/M_Z$ appearing in the IFI when $V=Z$ and vanishes when $V=\gamma$.\cite{Greco:1975ke,Greco:1975rm,Greco:1980mh} Specifically,
\begin{equation}
    \Delta B_4^Z=-2Q_I Q_F \frac{\alpha}{\pi} \ln\left(\frac{t}{u}\right)\ln\left(\frac{M_Z^2-iM_Z \Gamma_Z - s}{M_Z^2-iM_Z\Gamma_Z}\right), \qquad
    \Delta B_4^\gamma = 0.
\end{equation}
While not strictly a soft contribution, this correction is numerically significant, 
\begin{equation}
    \frac{\alpha}{\pi}\ln\left(\frac{\Gamma_Z}{M_Z}\right)\approx 0.008.
\end{equation}
This term is essential for obtaining the correct suppression of IFI at the $Z$ pole when combined with other CEEX contributions.

The integrals can be evaluated in the semi-soft limit, leading to a compact expression for the differential cross section at quark CM energy $\sqrt{\hat s}$ and photon energy fractions up to $v_{\rm max}$,
\begin{eqnarray}
    \frac{d\sigma}{d\Omega}({\hat s},v_{\rm max}) &=& \frac{3}{16}\sigma_0({\hat s}) \sum_{V,V'}\int_0^1
    dv dv' du du' \theta(v_{\rm max} - v - v' - u - u') e^{Y(p_1,p_2,q_1,q_2)}\nonumber\\
    &\times&\rho(\gamma_I, 1-v)\rho(\gamma_F, 1-v')
    \rho(\gamma_X, 1-u) \rho(\gamma_X, 1-u')\nonumber\\
    &\times&\frac{1}{4} {\rm Re}\sum_{\{\lambda\}} e^{\alpha\Delta B_4^V(s(1-v-u)} \Mmf^V_{\{\lambda\}}(s(v+u),t) \nonumber\\
    &\times&\left[e^{\alpha\Delta B_4^{V'}(s(1-v-u')} \Mmf^{V'}_{\{\lambda\}}(s(v+u'),t)\right]^*
\label{bigIntegral}\end{eqnarray}
where $Y(p_1,p_2,q_1,q_2)$ is the standard YFS\cite{Yennie:1961ad} form factor and radiative factor 
\begin{equation}
    \rho(\gamma,z)\equiv \frac{e^{-C_E \gamma}}{\Gamma(1+\gamma)}\gamma (1-z)^{\gamma-1}
    \label{rho}
\end{equation}
with Euler constant $C_E$, and 

\begin{equation}
    \gamma_I = Q_I^2 \frac{2\alpha}{\pi}\left[\ln\left(\frac{(p_1+p_2)^2}{m_I^2}\right)-1\right], \qquad
    \gamma_F = Q_F^2 \frac{2\alpha}{\pi}\left[\ln\left(\frac{(q_1+q_2)^2}{m_F^2}\right)-1\right], \nonumber
\end{equation}
\begin{equation}
    \gamma_X = Q_I Q_F \frac{2\alpha}{\pi}\ln\left(\frac{1-\cos\theta}{1+\cos\theta}\right) .
    \label{gammaFactors}
\end{equation}


KKhhFoam extrapolates this calculation to the entire phase space by replacing the additive constraint $(q_1+q_2)^2 = (p_1+p_2)^2(1-v-v'-u-u')$ by a multiplicitive ansatz 
\begin{equation}
    \frac{(q_1+q_2)^2}{(p_1+p_2)^2}=(1-v)(1-v')(1-u)(1-u')
\end{equation}
and upgrading the radiative factors $\rho$ in eq.\ (\ref{bigIntegral}) to order $\alpha^2$ following expressions from KKMChh. The complete order $\alpha$ virtual contributions are completed by adding the non-soft parts of the $\gamma\gamma$ and $\gamma Z$ box diagrams to the Born spin amplitudes, replacing $M(s,t)$ with 
\begin{equation}
    M(s,t) + M^{\gamma\gamma}(s,t,m_\gamma)+M^{\gamma Z}(s,t,m_\gamma)-2\alpha B_4 (s,t,m_\gamma)-\alpha\Delta B_4^Z(s,t).
\end{equation}
Electroweak corrections are included in the Born amplitudes via form factors calculated via Dizet 6.45, as in KKMChh. 

For given quark momenta and flavor, KKhhFoam must generate  
$v, v', u, u'$ and angles $\theta, \phi$ of the final lepton. Including the quark and antiquark momentum fractions and flavor, we
obtain a 9-dimensional integral, which is  evaluated by the Foam \cite{Jadach:1999sf,Jadach:2002kn} adaptive MC. Including parton distribution functions $f_q^h(x,{\hat s})$ for quark $q$ in hadron $h$ with momentum fraction $x$ and scale ${\hat s} = (p_1+q_1)^2 = sx_1 x_2$ (with $s = E_{\rm CM}^2$ in terms of the hadron CM energy) gives a cross section
\begin{equation}
    \sigma = \sum_q \int_0^1 dx_1 dx_2 f_q^{h_1}(x_1,{\hat s}) f_{\bar q}^{h_2}(x_2,{\hat s})\sigma_q({\hat s})
\end{equation}
with quark-level cross section $\sigma_q({\hat s})$ as described in eq.\ (\ref{bigIntegral}). 

In particular, the lepton invariant mass distribution takes the form 
\begin{eqnarray}
\frac{d\sigma}{dM_{l{\overline l}}} &=&
\frac{3\pi}{2}M_{l{\overline l}}\sum_q \int_{x_1 x_2 \ge s'/s}^1 dx_1 dx_2 
f_q^{h_1}(x_1,{\hat s})f_{\bar q}^{h_2}(x_2,{\hat s}) \sigma_0({\hat s})\int_{s'/s}^1 dz \rho(\gamma_I({\hat s}), z)
\nonumber\\
&\times& \int_{ww'\ge s'/s}^1 \frac{dwdw'}{ww'} \int_{-1}^1
d\cos\theta \rho(\gamma_X(\cos\theta),w) \rho(\gamma_X(\cos\theta), w') \rho\left(\gamma_F(s'), \frac{s'}{\hat{s} zww'}\right)\nonumber\\
&\times& \frac{1}{4}{\rm Re}\sum_{\{\lambda\}} e^{\alpha\Delta B_4^V({\overline s}w)} \Mmf_{\{\lambda\}}^V({\overline s}w,\cos\theta) \left[ e^{\alpha\Delta B_4^{V'}({\overline s}w')} \Mmf_{\{\lambda\}}^{V'}({\overline s}w',\cos\theta)\right]^*
\label{MassDist}
\end{eqnarray}
where we have defined $z = 1-v$, $z' = 1-v'$, $w = 1-u$, $w' = 1-u'$, and scales ${\hat s} = x_1 x_2 s$, ${\overline s} = z{\hat s}$, $s' = M_{l{\overline l}}^2 = zz'ww'{\hat s}$.
Note that the matrix element and conjugate matrix element are evaluated at different scales ${\overline s}w$ and ${\overline s}w'$, respectively. 

\section{Comparisons of KKhhFoam and KKMChh}
\label{sec:comparisons}
KKhhFoam can be compared to KKMChh for suitably inclusive observables, such as the dilepton mass distribution $M_{l{\overline l}}$ or the forward-backward asymmetry $A_{FB}$, which is defined using the scattering angle in the lepton CM frame, the Collins-Soper angle,\cite{Collins-Soper} which satisfies
\begin{equation}
    \cos\theta_{CS}={\rm sgn}(q_1^z+q_2^z) \frac{q_1^+ q_2^- - q_1^- q_2^+}{M_{l{\overline l}}\sqrt{(q_1^++q_2^+)(q_1^-+q_2^-)}}
    \label{csangle}
\end{equation}
with $q^\pm \equiv q^0\pm q^z$ (neglecting lepton masses). For the tests in this section, we consider proton scattering at $8$ TeV with muon final states in the range 60 GeV $< M_{l{\overline l}} <$ 150 GeV.  

Ideally, the ab-initio QED ISR of KKMC should be used with PDFs including pure QCD. However, real PDFs fit data that contains QCD at the input scale, which may or may not be (partially) removed before fitting. For the purpose of illustrating the size of the raw QED corrections, we consider NNPDF3.1-NLO PDFs without QED. QED ISR depends on the quark masses, as seen in the $\gamma_I$ factor in eq.\ \ref{gammaFactors} for the ISR radiator.
We will assume current quark masses for the light quarks, following the first implementation of QED corrections in PDFs, MRST2004\cite{MRST2004}. The PDG quark masses\cite{pdg:1} used here are $m_u = 2.2 {\rm MeV},$ $m_d = 4.7 {\rm MeV},$ $m_s = 150 {\rm Mev},$ $ m_c = 1.2 {\rm GeV},$ $m_b = 4.6 {\rm GeV}$. \cite{pdg:1}
For precision phenomenology, it would be necessary to investigate the degree to which QED contamination in the PDFs can be neglected, or develop a subtraction method to match CEEX ISR to a QED-corrected PDF. We will return to ISR questions in section \ref{sec:ISR}.

\begin{figure}[!ht]
  \centering
  \includegraphics[width=0.45\textwidth]{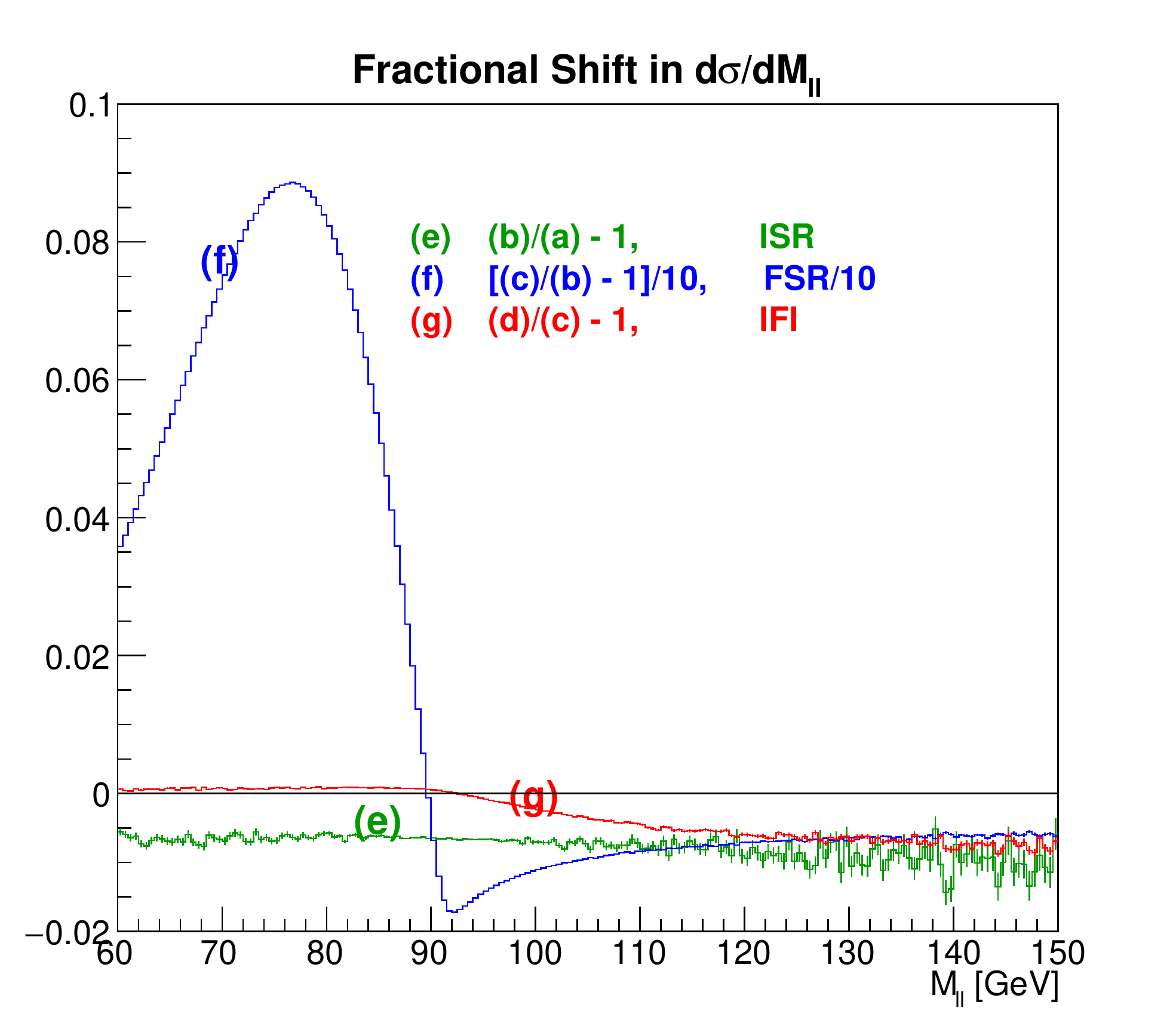}
  \includegraphics[width=0.45\textwidth]{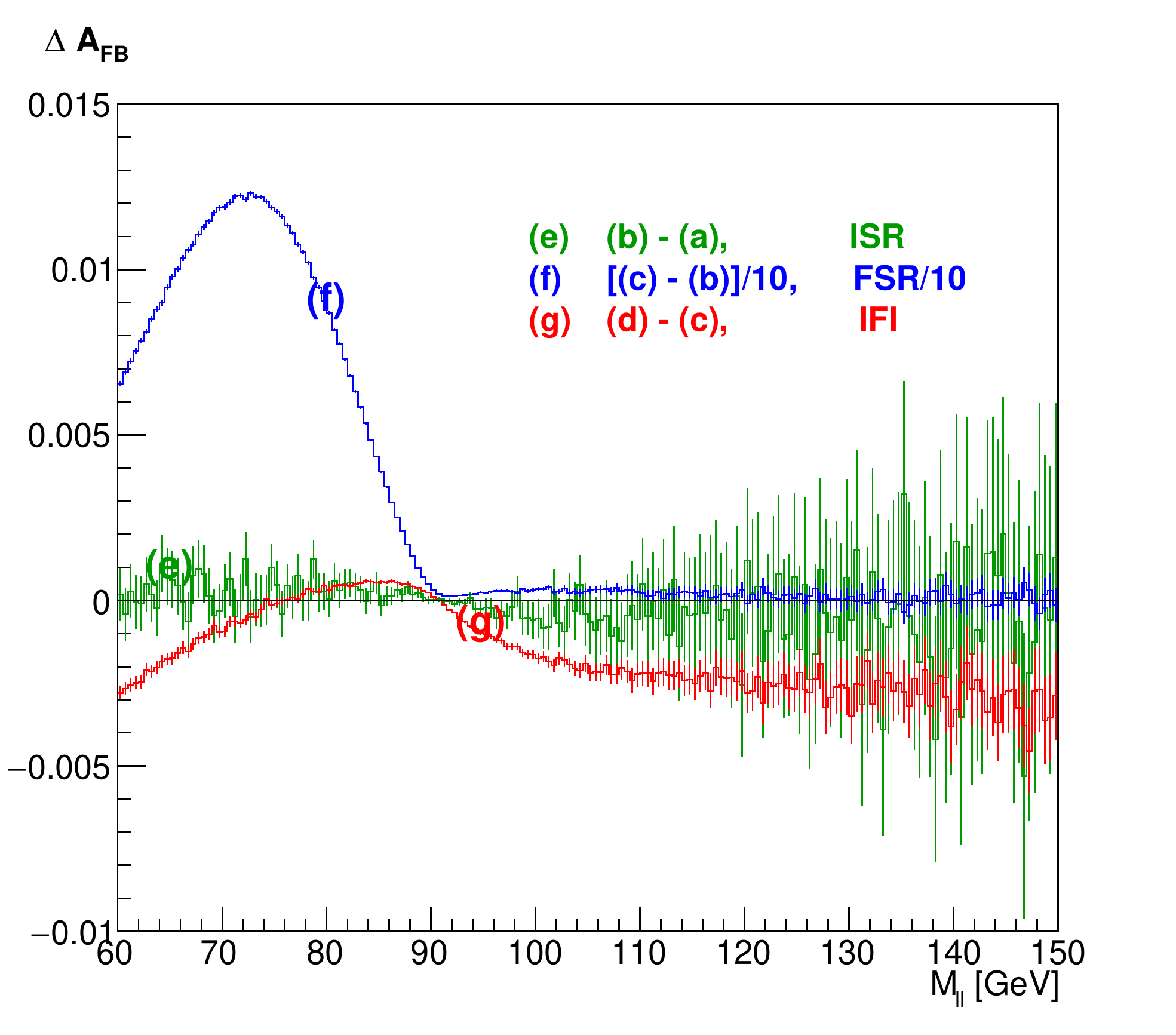}
  \caption{%
 This figure shows shows the effects of adding ISR (green), FSR (blue), and IFI (red) incrementally, with the FSR contribution divided by 10 so that it fits on the same scale. The figure on the left shows the fractional changes to the $M_{l{\overline l}}$ distribution for each, and the figure on the right shows the absolute changes in $A_{FB}$.
  }
  \label{fig:distributions1}
\end{figure}

Fig.\ \ref{fig:distributions1} shows the $M_{l{\overline l}}$ and $A_{FB}$ distributions for proton collisions at 8 TeV into muons, calculated in various ways, for a high-statistics KKMChh run with $23\times 10^9$ events ($10^{10}$ events for ISR only). Several different levels of photonic corrections are shown: without photonic corrections, with ISR, with ISR + FSR, and with ISR + FSR + IFI.  In the figure on the right, each of these additions is shown incrementally, in the order ISR, FSR, IFI. The photonic corrections are strongly dominated by FSR below $M_Z$. The FSR correction in the figure is divided by 10 so it fits on the scale of the ISR and IFI corrections. The downward shift by $\sim 0.5 - 1\%$ in the $M_{l{\overline l}}$ distribution due to FSR was seen as well in an ISR-only study with the same setup.\cite{Yost:2020jin} 

\begin{figure}[!ht]
  \centering
  \includegraphics[width=0.45\textwidth]{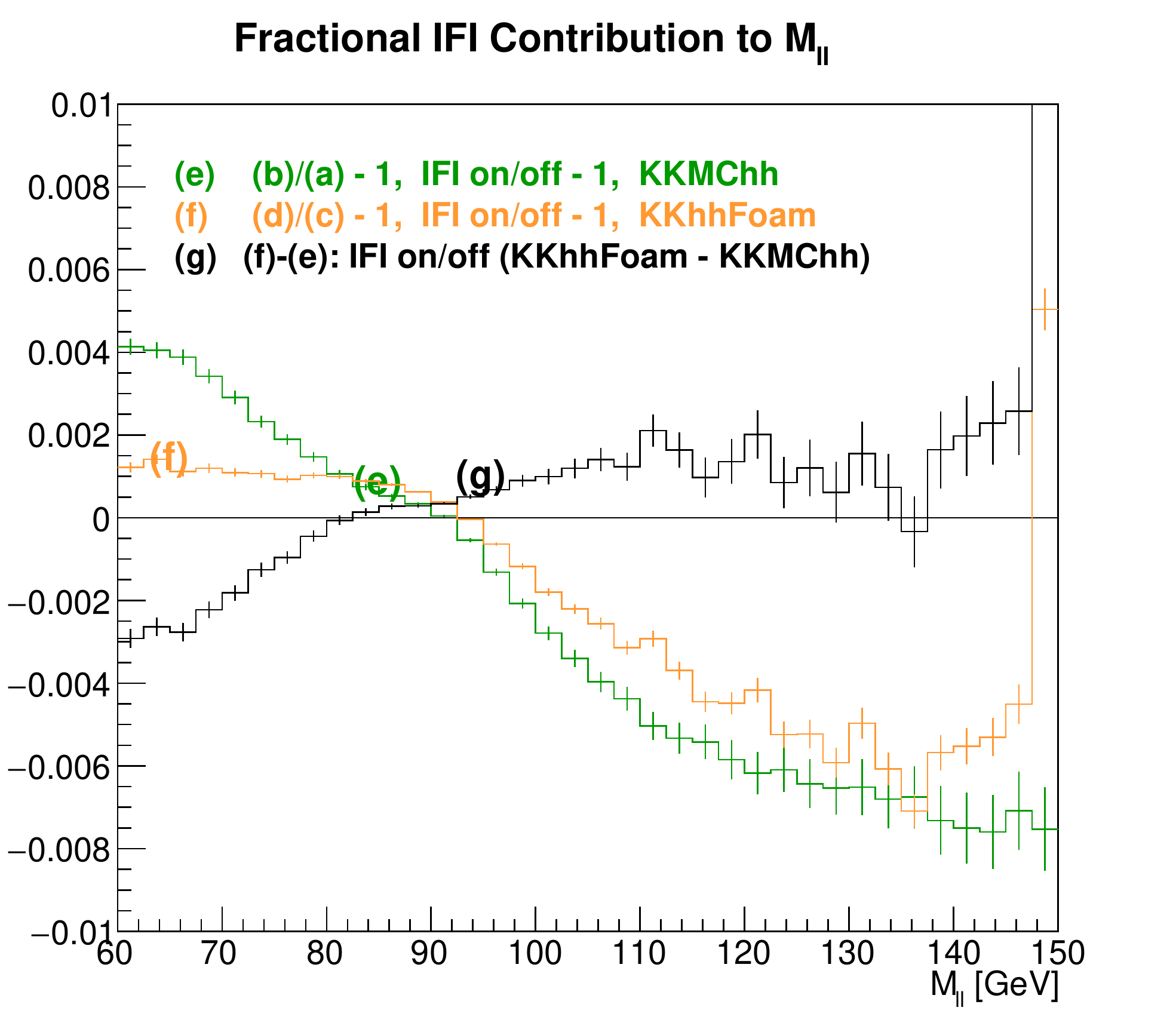}
  \includegraphics[width=0.45\textwidth]{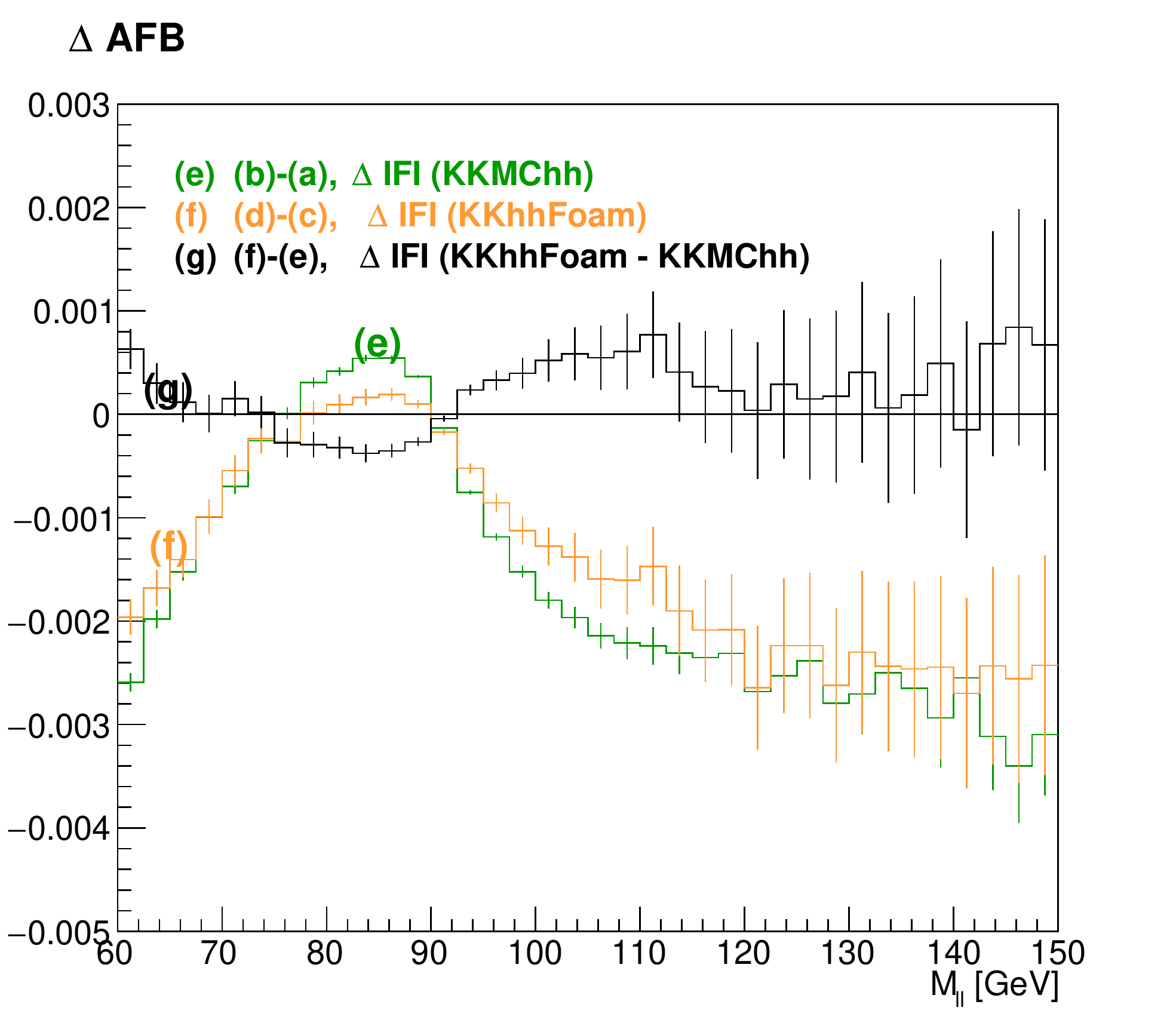}
  \caption{%
  Lepton mass distribution and charge asymmetry from \kkmchh\ and \kkhhfoam.
  The figure on the left shows the fractional contribution to $M_{l{\overline l}}$) of IFI for both \kkmchh\ and \kkhhfoam. The figure on the right shows the absolute contribution of IFI to $A_{FB}$. The black curves on the right are the differences between the \kkhhfoam\ and \kkmchh\ curves in each case. 
  }
  \label{fig:distributions2}
\end{figure}

Fig.\ \ref{fig:distributions2} shows comparisons of distributions calculated using KKMChh and KKhhFoam, focusing on initial-final interference. One of the key motivations for developing KKhhFoam, or the related $e^+e^-$ version \cite{Jadach:2018lwm}, was to have a cross-check of KKMChh with a comparable level of exponentiation. Such tests are important for a precision calculation of $A_{FB}$, which is important phenomenologically for measuring the electro-weak mixing angle. 

The difference in IFI corrections to $A_{FB}$ between KKMChh and KKhhFoam is less than $5\times 10^{-4}$ at energies below 100 GeV and generally less than $10^{-3}$, with decreasing statistics at higher energies. The fractional difference in the IFI correction to $d\sigma/dM_{l{\overline l}}$ is generally less than $0.2\%$. Some difference is expected because KKMChh includes complete $\Ocal(\alpha)$ hard photon corrections, which are missing in KKhhFoam. It is apparent that for $A_{FB}$, these hard corrections are quite small. The soft limit of KKMChh can be checked more precisely by including an artificial cutoff on the maximum photon energy. Such tests are presently in progress.

\section{Initial State Radiation}
\label{sec:ISR}

It is illuminating to rearrange the ISR radiators in eq. (\ref{bigIntegral}) or (\ref{MassDist}) using the fact that the basic radiator function, eq.\ (\ref{rho}) has a simple convolution property,
\begin{equation}
    \int_0^1 dv_1 dv_2 \delta (v-v_1-v_2)\rho(\gamma_1, 1-v_1), \rho(\gamma_2, 1-v_2) = \rho(\gamma_1+\gamma_2, 1-v).
    \label{convolution1}
\end{equation}
In the soft limit, with $z=1-v$, the constraint in eq.\ (\ref{convolution1}) can be replaced by $\delta(z-z_1 z_2)$, allowing the ISR radiator to be factorized in the form
\begin{equation}
    \rho(\gamma_I({\hat s},z)) = \int_0^1 dz_1 dz_2 \delta (z-z_1 z_2)\rho\left(\frac12 \gamma_I({\hat s}),z_1\right)
    \rho\left(\frac12 \gamma_I({\hat s}),z_2\right).
    \label{convolution2}
\end{equation}
Each ``half-radiator'' in eq.\ (\ref{convolution2}) can be combined with a PDF $f_q^h(x_i, {\hat s})$ in eq.\ (\ref{bigIntegral}) to make a QED-corrected PDF 
\begin{equation}
    F_q^h(x_i',{\overline s}) = \int_0^1 dx_i dz_i \delta (x_i'-z_ix_i)\rho\left(\frac12 \gamma_I({\hat s}), z_i\right) f_q^h(x_i, {\hat s})
    \label{QEDpdf}
\end{equation}
with ${\overline s} = z{\hat s} = sx_1'x_2'$. The ISR radiator in eq.\ (\ref{bigIntegral}) or (\ref{MassDist}) is then absorbed into QED-corrected PDFs with the replacement 
\begin{equation}
\int dx_1 dx_2 dz f_q^{h_1}(x_1,{\hat s}) f_{\overline q}^{h_2}(x_2,{\hat s})\rho(\gamma_I({\hat s}), z) = \int dx_1'dx_2' \delta({\overline s} - sx_1' x_2') F_q^{h_1}(x_1', {\overline s}) F_{\overline q}^{h_2}(x_2', {\overline s}).
\end{equation}

Ideally, the initial PDFs $f_q^h$ should model pure QCD, and $F_q^h$ would be a QED corrected version of this. However, real PDF sets always have some QED contamination in the input data. It is arguable that this QED contamination may have a negligible effect on calculations at $\sqrt{s}\sim M_Z$, but this assumption should be tested. There are a number of PDF sets available with QED evolution, beginning with MRST2004\cite{MRST2004} and including  NNPDF3.1-LuxQED \cite{NNPDF-LuxQED,LuxQED}, APFEL\cite{APFEL}, CT14QED\cite{CT14QED}, and MMHT2015qed\cite{MMHT2015qed}. A calculation of a suitably inclusive observable using KKMChh using a QCD PDF can be compared to the same calculation using a QED-corrected version of the same PDF. The two calculations should agree if the QED contamination in the original set is negligible. We will compare the $M_{l{\overline l}}$ distribution calculated for KKMChh with ISR only for NNPDF3.1nlo\cite{NNPDF3.1} and CT14nlo\cite{CT14} to the result with ISR off in KKMChh but the QED-corrected version of the same PDF set.

\begin{figure}[!ht]
  \centering
  \hbox{\kern-5pt
  \includegraphics[width=0.36\textwidth]{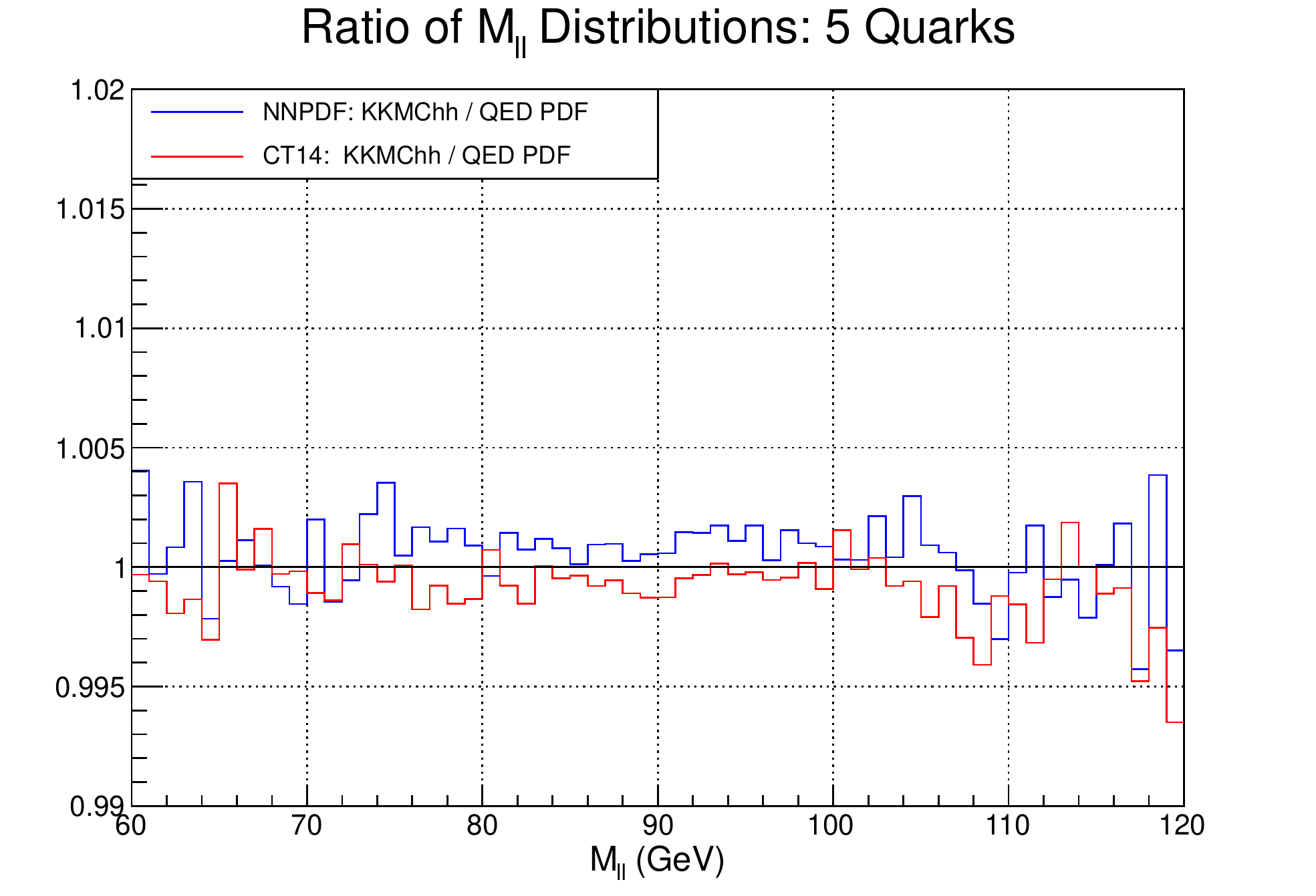}\kern-10pt
  \includegraphics[width=0.36\textwidth]{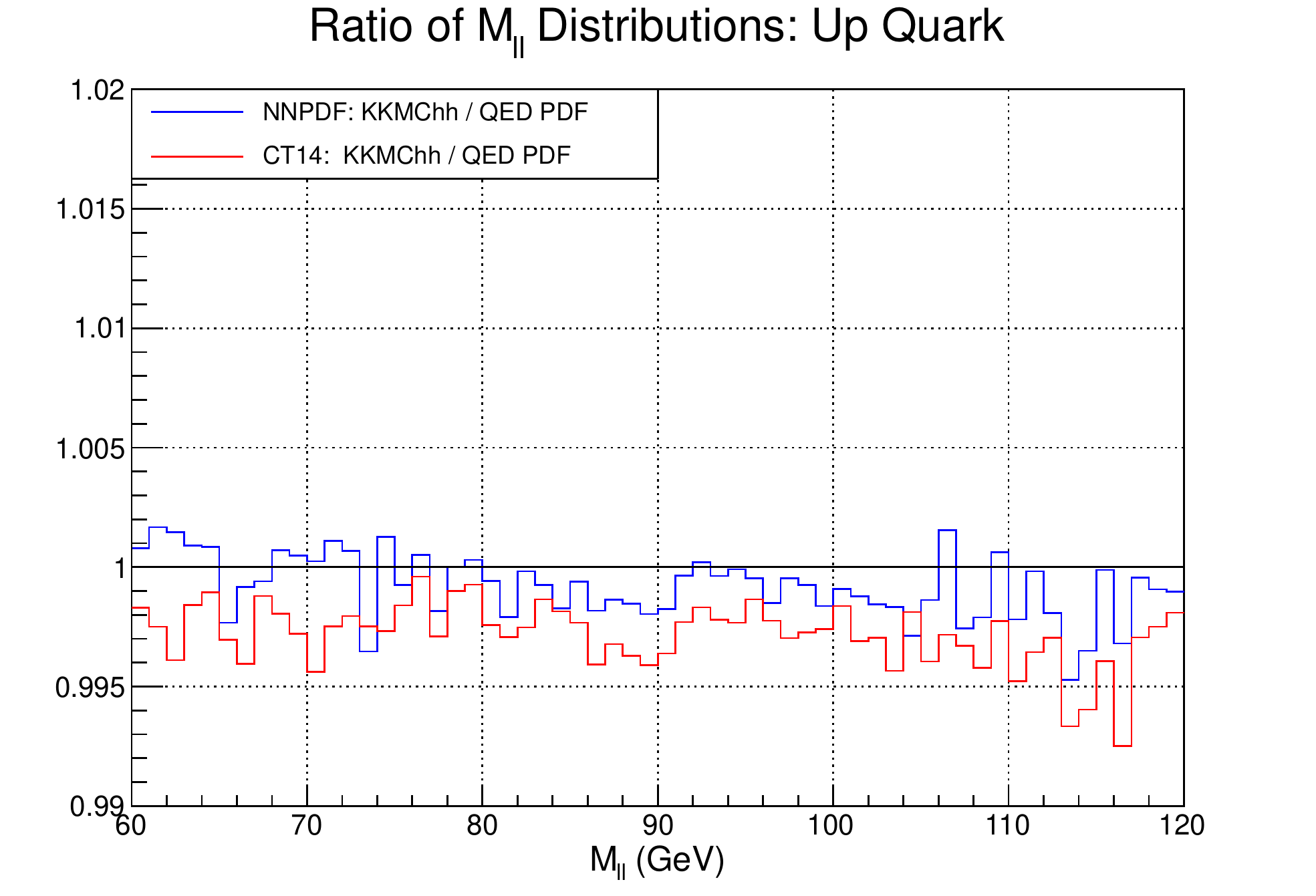}\kern-10pt
  \includegraphics[width=0.36\textwidth]{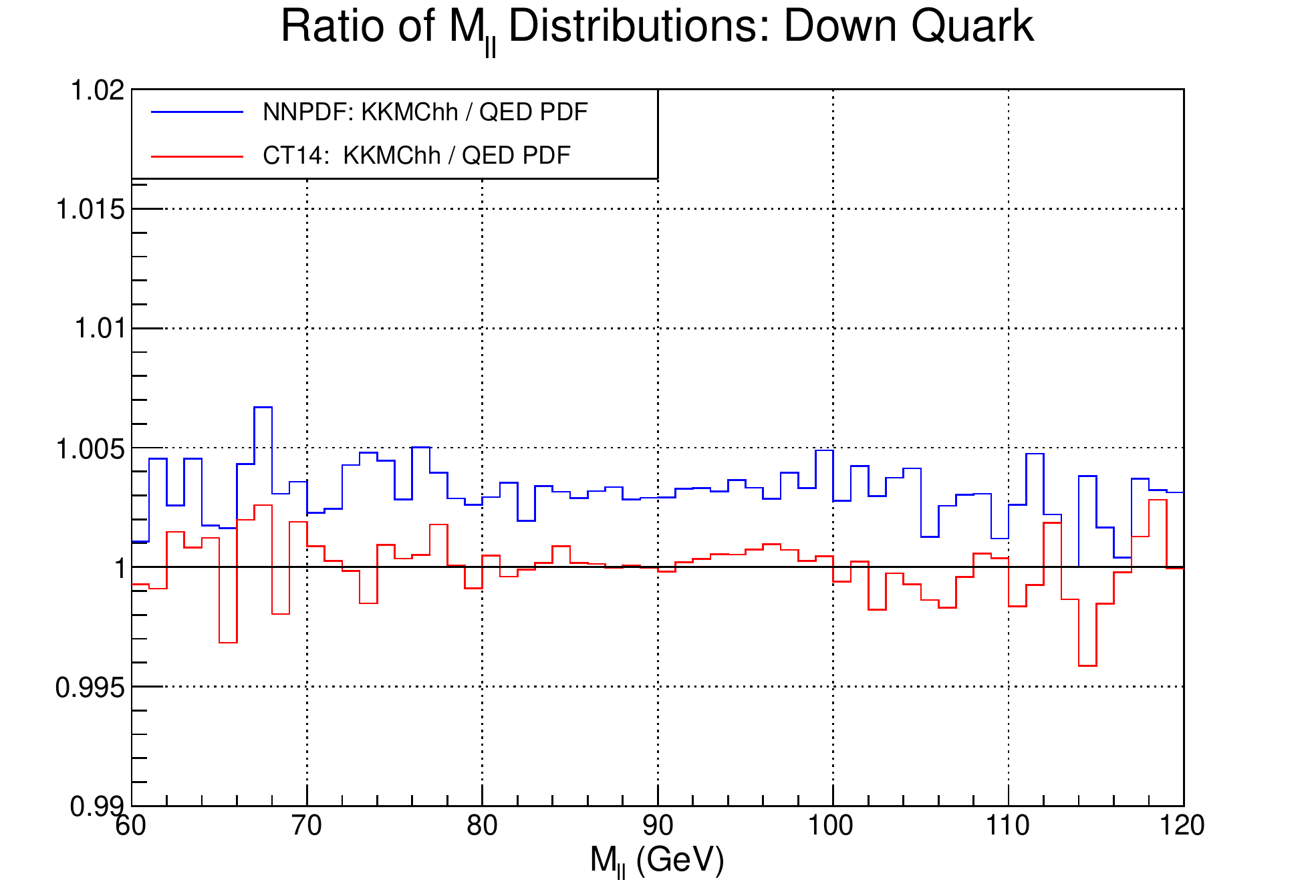}
  }
  \hbox{\kern-5pt
  \includegraphics[width=0.36\textwidth]{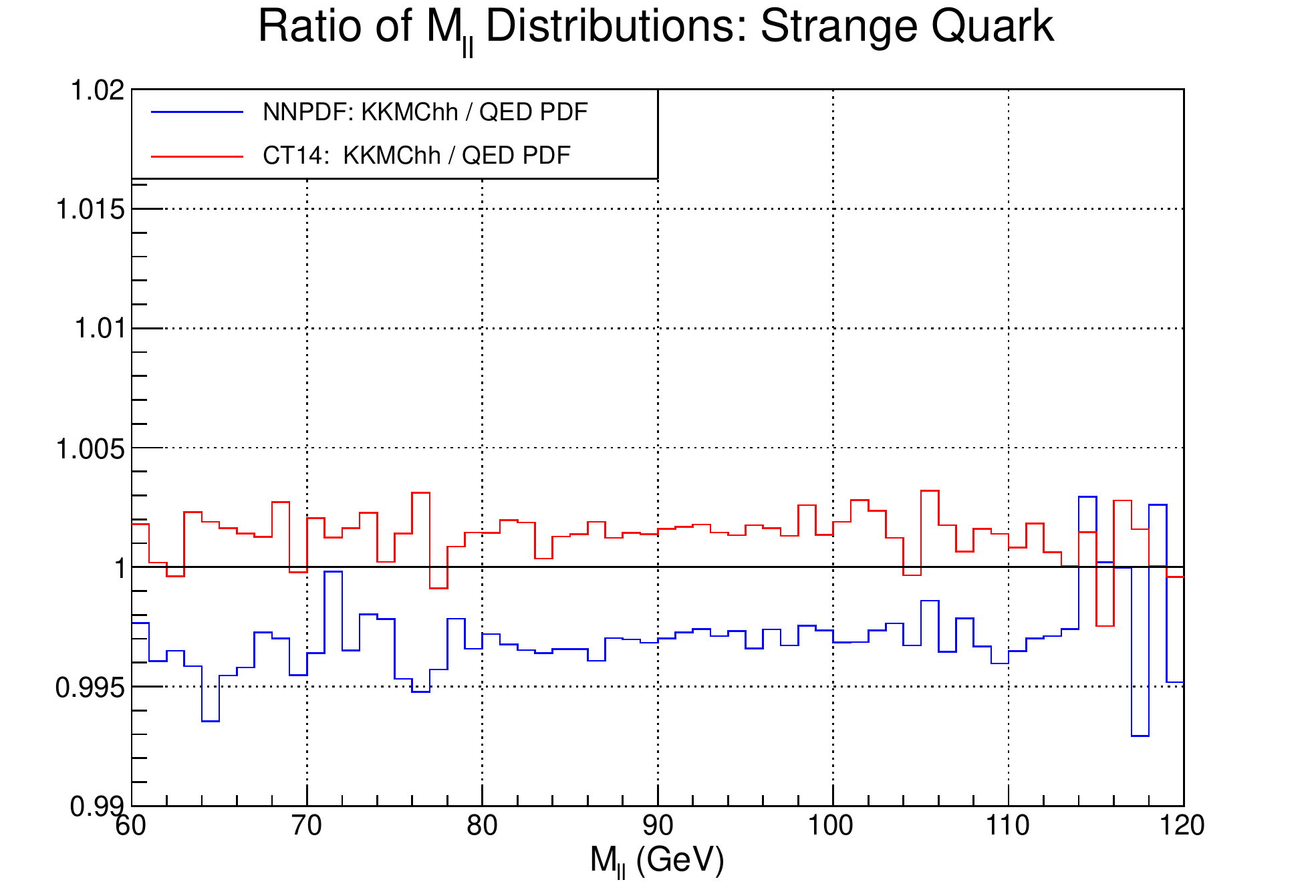}\kern-10pt
  \includegraphics[width=0.36\textwidth]{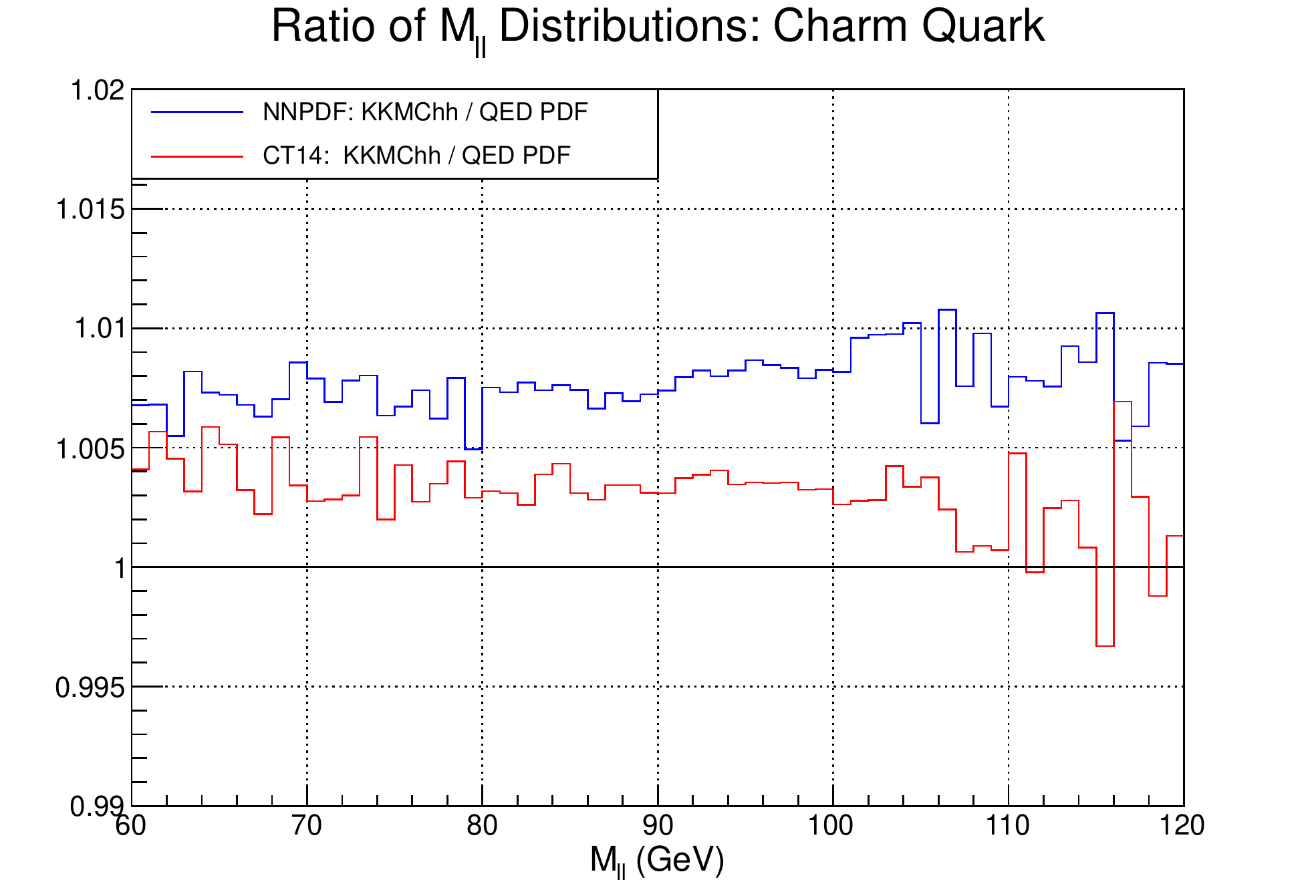}\kern-10pt
  \includegraphics[width=0.36\textwidth]{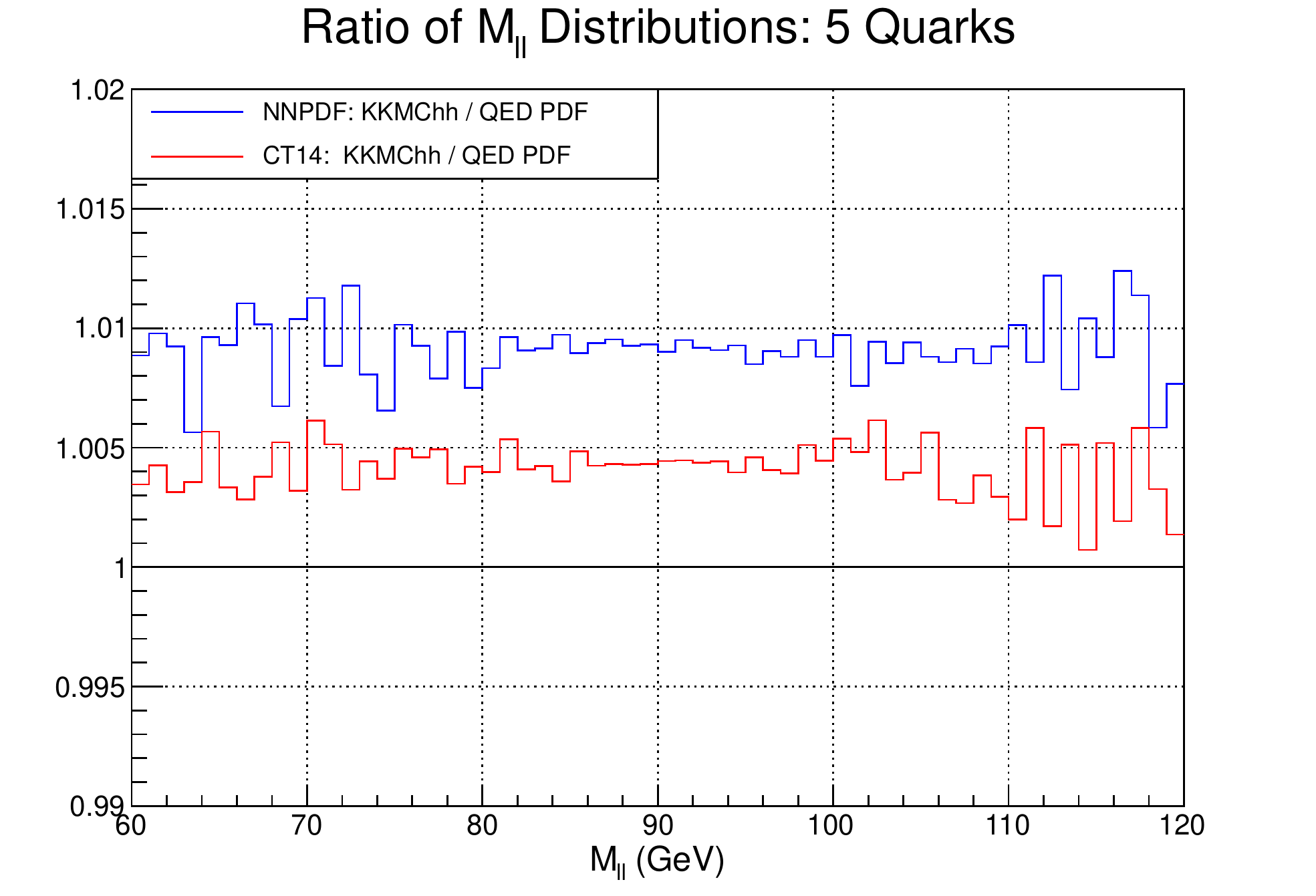}
  }
  \caption{%
   These figures show the ratio of a QED-corrected PDF set to a standard one for two sets: NNPDF3.1 in blue, and CT14 in red. In each case, the relative KKMChh ISR correction is shown for each standard PDF set: NNPDF3.1 in green, and CT14 in orange. The first figure includes all five quarks, while the remaining figures show the individual up, down, strange, charm, and bottom quarks.
  }
  \label{fig:ISR}
\end{figure}

Fig.\ \ref{fig:ISR} show the ratio of the KKMC-hh dimuon mass distribution with ISR on and a standard PDF set to the same distribution calculated with ISR off and a QED-corrected PDF set, for $10^9$ muon events in $8$ TeV proton collisions, as in the previous section. The blue histograms are for NNPDF3.1nlo, and the red histograms are for CT14nlo. NNPDF3.1nlo and CT14nlo. The result for all five quarks is shown in the first plot, and the subsequent plots show the results for the up, down, strange, charm, and bottom quarks invidually. Both NNPDF and CT14 give ratios that agree with 1 to within $\pm 0.001$, roughly the size of the statistical errors and within the difference between the ratios for the two PDFs. The complete result is strongly dominated by the up quark, which shows a comparable level of agreement. For the down quark, CT14 gives a ratio consistent with 1, but the result with KKMChh ISR is about $0.5\%$ higher than the result with NNPDF3.1-LuxQED. The pattern of KKMChh matching CT14 more closely than NNPDF continues with the heavy quarks, but in all cases, the agreement with KKMChh is at a level comparable to the agreement between the different QED-corrected PDFs. This suggests that the effect of any QED contamination in the original PDFs is likely to have a negligible influence on KKMChh calculations.

Although the agreement between the QED ISR added by KKMChh and by a QED-corrected NNPDF or CT14 PDF is good, the practice of fitting data containing QED to a PDF with pure QCD evolution does not provide an ideal starting point. A firmer starting point may be to start with a PDF set that acknowledges both QCD and QED evolution, and removing ISR via backward evolution of the $\rho$ factors, going from $F_q^h$ to $f_q^h$ in eq.\ (\ref{QEDpdf}), pruning the QED from the PDF before putting it back in KKMChh. This procedure is presently being tested, and appears to be promising, with the added bonus that the quark masses in the ISR radiators will cancel between the forward and backward evolution. Details will be reported soon. 

\section{Conclusion}
KKMChh has been newly reprogrammed in C++, facilitating integration with modern showers such as HERWIG7 and the introduction of NLO QCD, which will be an important next step. The semi-analytical program KKhhFoam provides a useful cross-check of KKMChh in the semi-soft limit, as well as a way to better understand the structure of its amplitude-level soft photon exponentiation (CEEX). While KKhhFoam is less complete than KKMChh, it has has the benefit of allowing ISR, FSR and IFI to be switched on independently. 

Finding the best approach to integrating KKMChh's ISR with real PDF sets is an ongoing project. It appears that combining KKMChh with a standard QCD PDF is likely to be adequate for current phenomenological purposes, but a new alternative, suggested by KKhhFoam, is to use backward evolution via the exponentiated ISR radiator to remove QED ISR at a selected factorization scale, allowing KKMChh to run with QED-corrected PDF sets. This approach is more satisfying theoretically, since it starts with PDFs that acknowledge a mixture of QCD and QED evolution. It also follows the more familiar approach of factorizing collinear ISR into the PDFs. We expect to report on this development soon. Whether this approach is better phenomenologically will depend on the precision of the QED modeling in the PDFs.      

\section*{Acknowledgements}
We acknowledge computing support from the Institute of Nuclear Physics IFJ-PAN, Krakow, and The Citadel. 
S.J. acknowledges funding from the European Union’s Horizon 2020 research and innovation programme under under grant agreement No 951754 and support of the National Science Centre, Poland, Grant No. 2019/34/E/ST2/00457.


\bibliography{radcor2021-kkmchh.bib}
\nolinenumbers
\end{document}